\documentclass{article}
\usepackage{graphicx}
\usepackage{amssymb,amsmath}
\usepackage{color}
\usepackage{geometry}
\usepackage{natbib}

\begin{document}
\begin{center}
\large
\textbf{Holographic interferometry study of the dissolution and diffusion of
gypsum in water} \\
\vspace{0.5cm}
\normalsize
\textsc{Jean Colombani\footnote{Author to whom correspondence should be addressed
(Jean.Colombani@lpmcn.univ-lyon1.fr).}} and \textsc{Jacques Bert}\\
\small
Laboratoire de Physique de la Mati\`ere Condens\'ee et Nanostructures,
Universit\'e Claude Bernard Lyon 1;\\
CNRS, UMR 5586,
Domaine scientifique de la Doua,
F-69622 Villeurbanne cedex, France\\
\normalsize
\vspace{1cm}
Revised version - 29 November 2006


\begin{minipage}{15cm}\textbf{Abstract---} 
We have performed holographic interferometry measurements of the dissolution of
the (010) plane of
a cleaved gypsum single crystal in pure water.
These experiments have provided the value of the dissolution rate
constant $k$ of gypsum in water and the value of the interdiffusion coefficient $D$ of its aqueous
species in water.
$D$ is $1.0\times10^{-9}$ m$^2$ s$^{-1}$, a value close to the
theoretical value generally used in dissolution studies.
$k$ is $4\times10^{-5}$ mol m$^{-2}$ s$^{-1}$.
It directly characterizes the microscopic transfer rate
at the solid-liquid interface,
and is not an averaged value deduced from quantities measured far from the
surface as in macroscopic dissolution experiments.
It is found to be two times lower than the value
obtained from macroscopic experiments.
\end{minipage}
\end{center}


\section{INTRODUCTION}

Pressure Solution Creep (PSC) constitutes a major plastic strain process
of immersed solids under low stress.
A crystal wetted by a solvent (generally water) and saturated by its components
is at chemical equilibrium.
But if that crystal is then
submitted to an external stress, it experiences dissolution.
This originates in the chemical potential change of the solid induced by the
stress.
The dissolved species then diffuse away from the high stress region
and precipitate in a stress-free zone.
This mechanism is known to play a chief role in the upper crust and
contributes, for instance, to upper crust deformation
and to the diagenesis of sedimentary rocks.

The study of PSC in gypsum is of double interest.
First it controls the upper crust strength in numerous geological situations
because of its high ductility \citep{deMeer97}.
Additionally, PSC in gypsum also likely plays a role in the high ductility and low strength
of plaster, which is essentially made of gypsum needles,
in the presence of moisture \citep{Chappuis}.

To understand the mechanisms involved in PSC in gypsum,
precise values of the involved quantitative parameters are needed.
Among them, the diffusion coefficient and the dissolution rate constant are
primordial.
Furthermore, these values are also of interest in other fields.
Wide areas of gypsum karsts exist world-wide
The instability of these karsts and potential collapse undermines the importance
of understanding karst evolution and subsequently, the importance
of knowing the precise values involved
in the water-gypsum interaction \citep{Jeschke}.
The presence of the dissolved species of gypsum in water influences the dissolution
of minerals containing pollutants (especially in uranium mines).
Hence the study of such contamination also necessitates values regarding gypsum dissolution
\citep{Kuechler}.
The existence of a large quantity of dissolved mineral may also alter the quality
of the drinking water \citep{Raines}.
Lastly, knowledge of gypsum dissolution/precipitation behaviour is required
in the oil and gas industry where gypsum is a common annoying "scale" mineral
\citep{Raju}.

Surprisingly, no diffusion coefficient measurements of gypsum in water are
available in the literature \citep{Lobo,Zaytsev}.
Recent studies of gypsum dissolution either estimate the
interdiffusion coefficient of the aqueous components \citep{Raines}
or make use of a numerical value computed in 1971 \citep{Barton} with a Nernst-Hartley
equation from the tracer diffusivity of the ions \citep{Jeschke}.
Classical dissolution measurements typically
use rotating-disk setups, batch-experiments or column-experiments
with powder or crystal samples.
In these apparatuses, the concentration during dissolution is monitored by
techniques such as
titrimetry, conductimetry, atomic absorption spectroscopy,
mass spectrometry or colorimetry.
During these experiments, water is usually flowing and the overall measured
concentration
stems from a combination of diffusion, forced convection and dissolution.
If empirical equations always enable one to describe the measurements, the link
with pure phenomena is not straightforward and needs 
accurate knowledge of the surface morphology and of the transport
properties of the system to be reliable \citep{Jeschke02b}.
To overcome these experimental limitations, we have carried out local, instead
of global, experiments of gypsum dissolution in water
with an alternative technique, holographic interferometry.
This method allows simultaneous determination of the dissolution
rate constant of gypsum in water at the dissolving
interface and of the diffusion coefficient of its components
in bulk water.
The fundamental difference between our measurements and standard
ones lies in the fact that we look directly at the surface behavior instead of
deducing this behavior from quantities measured far from the surface.

\section{EXPERIMENTS}

Historically,
real-time holographic interferometry has been recognized as a valuable tool in
solution chemistry \citep{Knox}.
One of its advantages is that defects along the optical
path (except in the working cell) compensate between the two light expositions
of the hologram (see below), thus requiring less demanding setups than classical
interferometry devices \citep{Colombani06}.
Furthermore the observation of the two-dimensional concentration field in the
liquid enables the
identification of any non-diffusive fluxes (natural convection, gravitational
instability, \dots), thus providing a guarantee of the reliability of the
measurements \citep{Colombani98}.

\begin{figure}
\begin{center}
\includegraphics[width=\textwidth]{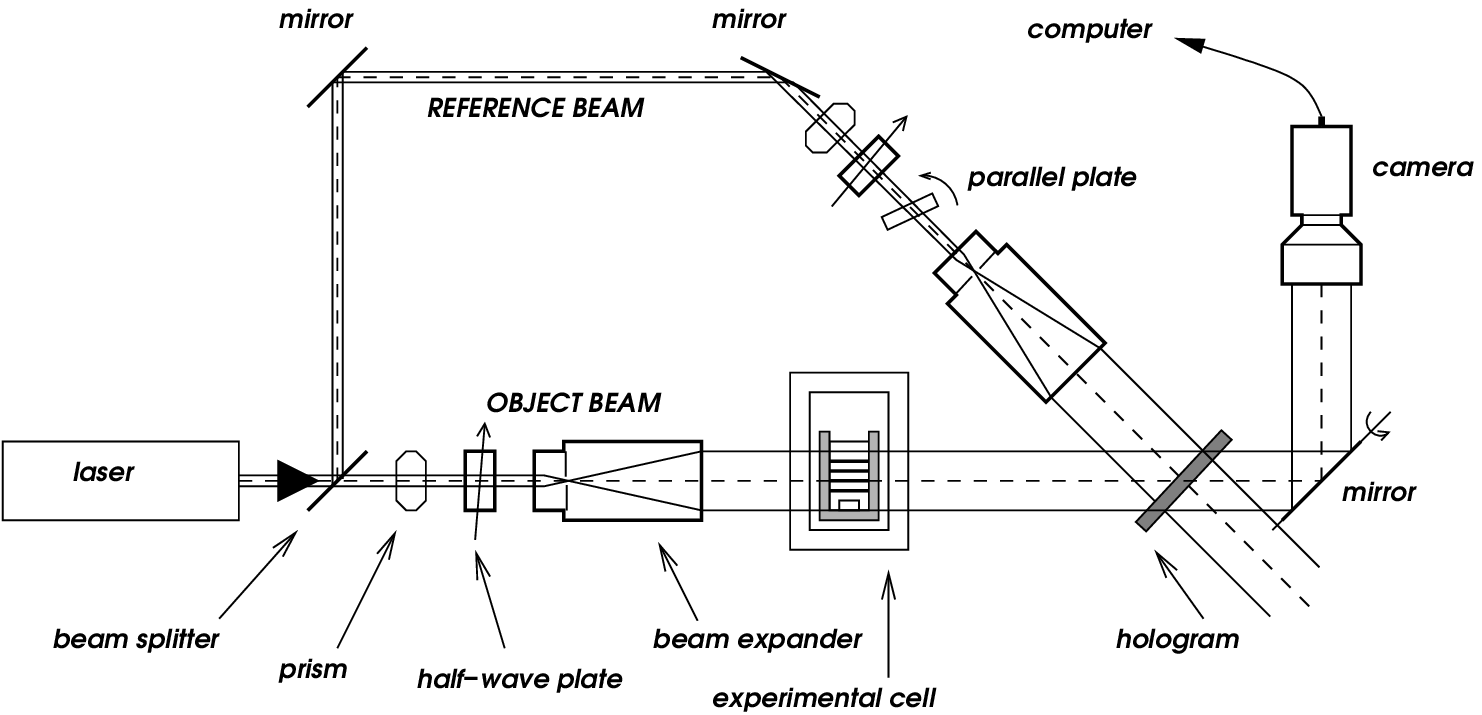}
\caption[Holographic setup:  DPSS laser, beam splitter,
Glan-Taylor prism, half-wave plate with adjustable polarization direction,
beam expander comprising
a microscope objective, a pinhole and a converging lens,
thermostated optical cell, holographic plate, mirrors, 
parallel plate on a rotating mount, camera.]{}
\label{optic}
\end{center}
\end{figure}

Our setup is designed as follows (see Fig. \ref{optic}).
The beam of a Diode Pumped Solid State laser ($\lambda=$ 532 nm wavelength)
is divided into a reference beam and an object
beam by a 50/50 beam splitter.
The two beams cross a half-wave plate and a Glan-Taylor prism
(vertically polarizing), which act as
a polarizer/analyzer system controlling the intensity of the beam.
The two beams are both expanded and spatially filtered by
a microscope objective/pinhole/convergent lens set.
A parallel plate installed on a rotating mount is
inserted between the pinhole and the convergent lens of the reference
beam.
This device allows the removal of eventual parasitic fringes on the initial
interferogram.
The two beams interfere on the holographic plate, their
polarization vectors being parallel and colinear to the plate.
The entire setup is located on a vibration-damping structure.

Air motion in the laboratory causes a random change with time
of the path length along the
beams, creating parasitic vibrations of the fringes.
To reduce this source of uncertainty, the entire optical table is covered with a
closed shelter and the real-time optical data acquisition is
performed in a contiguous room instead of in the laboratory.

The primary interest of holography lies in the fact that
the interference pattern of the object and the reference
beams contains the amplitude (as in classical photography)
and the phase of the object beam.
In this way, the amplitude and phase of
the object at time $t_0$ are recorded on the hologram.
Therefore, when both the object and the hologram (containing the
memory of the object at time $t_0$) are enlightened at time $t$ by the laser,
the phase difference $\Delta\phi$ of the object between times $t_0$ and $t$
are visualized through $N=\Delta\phi/(2\pi)$ interference fringes (interferograms).
The interferograms are visualized and recorded with a
Charge-Coupled Device camera, a monitor and image acquisition software.

This device, thus, gives access to any evolution of the phase
$\Delta\phi=2\pi e\Delta n/\lambda$ ($e$ path length in the optical
cell) and consequently to any evolution
of the refractive index $\Delta n$
of the liquid (all other optical components being kept unchanged).
Therefore, the change of solute molality
$\Delta m=\Delta n/(\partial n/\partial m)=N\lambda/(e(\partial n/\partial m))$
in the solution can be registered
($\partial n/\partial m$ is the derivative of the index of refraction $n$
of the solution with the solute molality $m$).

The crystal/solution system is located in an optical cell inside
a copper structure with circulating thermostated water.
The temperature of the water bath is regulated
by a PID regulator, which probes the temperature with a platinum thermistor
inserted into the copper structure.
Indeed a temperature variation may be the source of a solubility, diffusivity
and dissolution rate constant change.
Therefore, constant temperature must be guaranteed ($\pm$0.01 K
in our experiments).

\begin{figure}
\begin{center}
\includegraphics[width=.8\linewidth]{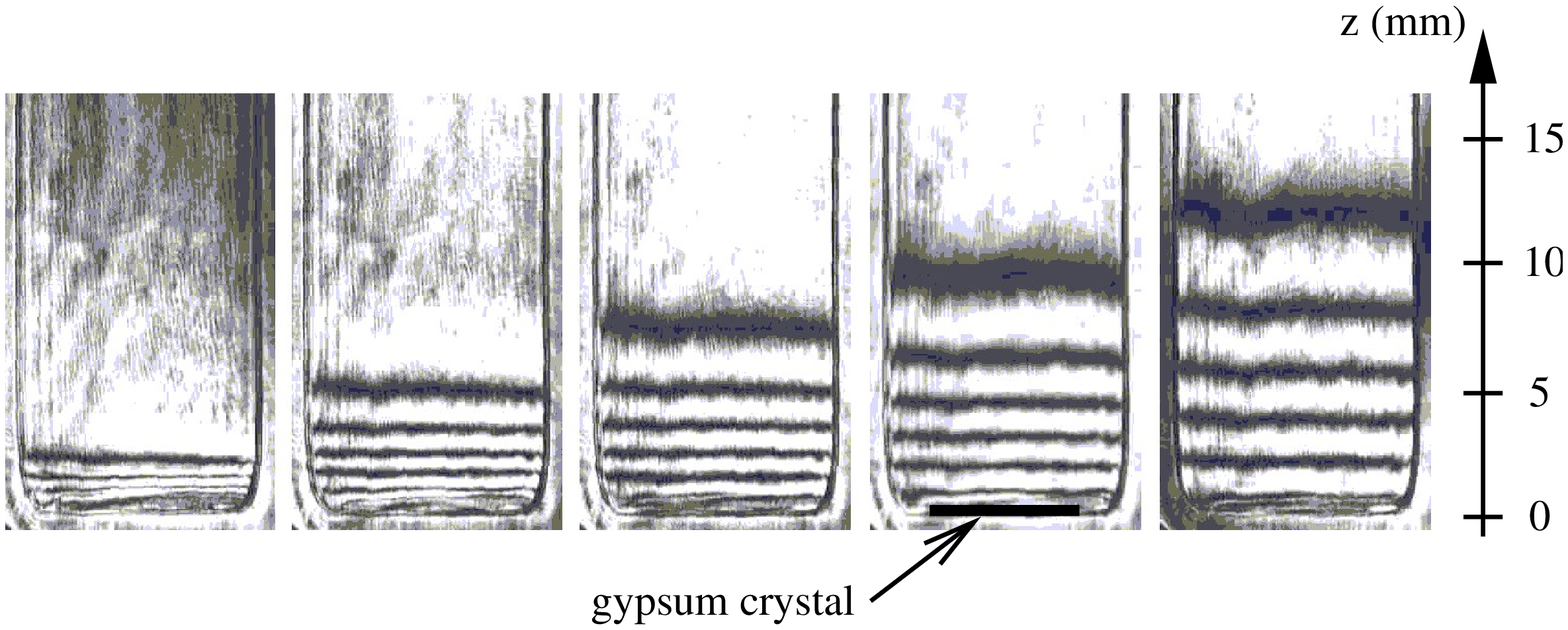}
\caption[Interferograms during the
dissolution of a gypsum single crystal in pure water
9, 60, 120, 180 and 400 minutes
after the beginning of the experiment.
A black bar indicates the position of the crystal in one of the
interferograms.]{}
\label{interf}
\end{center}
\end{figure}

\begin{figure}
\begin{center}
\includegraphics[width=.6\linewidth,angle=-90]{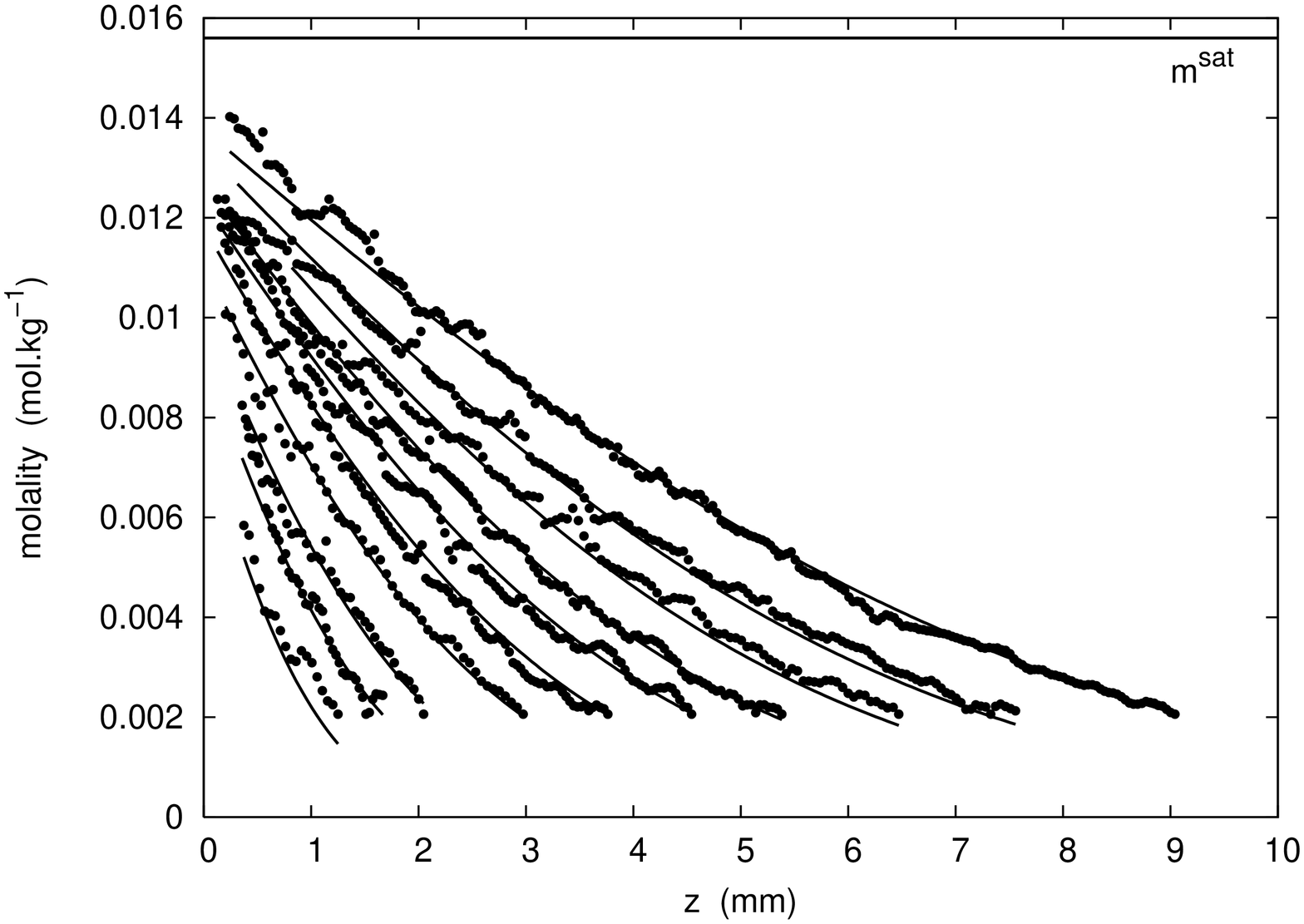}
\caption[Evolution of the molality with vertical position during the
experiment of Fig. \ref{interf}, 9, 20, 30, 60, 75, 105, 120,
180, 240 and 360 min.
after the beginning of the experiment.
Circles are for experimental points and lines for theoretical fit.
The horizontal line is the solubility.]{}
\label{sscont}
\end{center}
\end{figure}

An experiment proceeds as follows:
The optical cell is filled with ultrapure water.
The liquid is left in the copper structure for a few hours
in order to obtain a homogeneous temperature in the system.
Subsequently, a reference hologram is taken.
Then a cleaved gypsum single crystal (CaSO$_4$,2H$_2$O) from the Mazan mine
(Vaucluse, France) of rough dimensions 5$\times$5$\times$1 mm$^3$
is introduced into the cell, this time being considered as the time origin
of the reaction.
Interferograms are then recorded periodically (see Fig. \ref{interf}).
The white homogeneous zone above the fringes
in the first interferograms
is considered to be pure water (the inhomogeneous grey zone is a parasitic fringe).
Indeed, no change of color is observable in this part of the cell
compared to the reference hologram, so the phase has remained unchanged and we may infer
that no discernible gypsum has migrated yet to this zone.
Proceeding downwards, an initial black fringe is encountered, revealing a change of $\pi$
in the phase of the transmitted beam, compared to the flat-colored zone just above.
Hence, we can conclude that the solute molality has increased at this point
until it has reached a value for which the resulting change of phase is $\pi$.
This corresponds to a molality change $\Delta m=\lambda/(2e(\partial n/\partial m))$
(cf. above).
The liquid just above the fringe has been identified as pure water therefore the
molality all along this fringe is merely $m=\Delta m$.
The adjacent white fringe, immediately below, corresponds to a $\pi$ change of the phase
compared to the black fringe, and a $2\pi$ evolution compared to the flat-colored zone.
The molality at this level is then $m=2\Delta m$.
The adjacent black fringe, just below, corresponds to a $\pi$ change of the phase
compared to the white fringe, and a $3\pi$ evolution compared to the flat-colored zone.
The molality at this level is then $m=3\Delta m$.
The molality in the whole cell is then gradually reconstructed
(see Fig. \ref{sscont}).

The limiting factor of the spatial resolution of the interferogram is
the number of pixels in the camera sensor.
The larger the pixel number, the smaller the area depicted by one pixel.
The size of the chip of our camera is 752$\times$582 pixels and our resolution 40
$\mu$m.
This value is small enough to provide us with a correct discretization of the molality
in the cell.

\section{DATA ANALYSIS}

The investigated chemical reaction can be written as :
\begin{equation}
\text{CaSO}_4\cdot 2\text{H}_2\text{O(s)}\quad \leftrightharpoons\quad
\text{Ca}^{2+}\text{(aq)} + \text{SO}_4^{2-}\text{(aq)} + 2\text{H}_2\text{O(l)}
\label{reaction}
\end{equation}
It takes place at the lower end of our cell.
Considering the symmetry of the experiment, it can be reduced to a
one-dimensional diffusion problem along the vertical coordinate $z$.
In the absence of other added salts, the two aqueous species are present at each $z$ value
with the same molality to preserve local electroneutrality.
Therefore, $m_{\text{Ca}^{2+}}(z,t)=m_{\text{SO}_4^{2-}}(z,t)$, $m_{\text{Ca}^{2+}}(z,t)$
and $m_{\text{SO}_4^{2-}}(z,t)$ being the molalities of Ca$^{2+}$(aq) and
SO$_4^{2-}$(aq), respectively, at position $z$ and time $t$.
Our investigation method is solely sensitive to the change of refractive
index $n$, regardless of the species causing this change.
In fact, both ions induce the modification of $n$, plus a possible contribution of
ion pairs and impurities.
Therefore, we experimentally access an effective molality of dissolved gypsum
$m(z,t)\approx m_{\text{Ca}^{2+}}(z,t)=m_{\text{SO}_4^{2-}}(z,t)$.

To evaluate $m$ along the vertical coordinate as
a function of time, Fick's second law is used.
if the diffusion coefficient is considered as
constant in our molality range (see Section \ref{Results}), this law reads:
\begin{equation}
\left(\frac{\partial m}{\partial t}\right)_z=
D\left(\frac{\partial^2 m}{\partial z^2}\right)_t
\label{Fick2}
\end{equation}
with $D$ as the interdiffusion coefficient of the solution,
and must be solved in space and time with the following boundary conditions
(all the symbols used are summed up in Table 1):
\begin{itemize}

\item Mass balance at the dissolving interface implies
the amount of gypsum leaving the solid and entering the liquid to being equal:

\begin{equation}
F_{\text{dissolution}}(0,t)=F_{\text{diffusion}}(0,t)
\label{massbal}
\end{equation}
where $F$ is the flow rate and the origin of coordinate is the geometric
solid-liquid interface, considered as fixed (actually moving of less than one
pixel in the course of a typical experiment).

The expression of the dissolution flow rate is:
\begin{equation}
F_{\text{dissolution}}(0,t)=\frac{d\xi}{dt}
\label{dissol}
\end{equation}
with $\xi$ the advancement variable (amount of transformed reagent).
If we follow a transition states theory \citep{Shiraki}, the rate of advancement
$d\xi/dt$ takes the form:
\begin{equation}
\frac{d\xi}{dt}=k s_r a_{sol}\left[1-\text{exp}{(-\frac{\mathcal{N}A}{R T})}\right]
\label{trans}
\end{equation}
where $k$ is the dissolution rate constant of gypsum in water,
$s_r$ is the total dissolving gypsum-water interface, $a_{sol}$ is the
activity of the reagent (conventionally taken as one for a solid),
$A$ is the chemical affinity of the reaction,
$R$ is the gas constant, $T$ is the absolute temperature and $\mathcal{N}$
is a constant.
The chemical affinity of dissolution, in other words the change of
Gibbs free energy when gypsum changes its thermodynamic state
from solid to dissolved, is $A=-RT\ln{\Omega}$.
In this expression, $\Omega$ stands for the supersaturation:
\begin{equation}
\Omega=\frac{a_{\text{Ca}^{2+}}a_{\text{S0}_4^{2-}}a^2_{\text{H}_2\text{O}}}{K_{sp}}.
\end{equation}
$a_X$ is the activity of the subscripted aqueous species and $K_{sp}$ is the solubility
product
of the reaction of Eq. \ref{reaction}\footnote{Our standard state is
characterized by
a unit activity for pure gypsum (solid reagent) and pure water (solvent) at any
temperature and pressure.
For the aqueous species of the solute, it is a unit activity in a 1 mol kg$^{-1}$
solution representing an infinitely dilute solution, for any temperature and pressure.}.
The solubility product can be written as:
$K_{sp}=a^{\text{sat}}_{\text{Ca}^{2+}}a^{\text{sat}}_{\text{S0}_4^{2-}}
(a^{\text{sat}}_{\text{H}_2\text{O}})^2$, where $a^{\text{sat}}_X$ represents
the activity of the subscripted species at chemical equilibrium.
As the solubility of gypsum is small,
the activity of the solvent $a_{\text{H}_2\text{O}}$ is always considered to be
equal to one.
If we introduce the mean ionic activity coefficient $\gamma_{\pm}$, geometric
mean of the two single ion activity coefficients, the supersaturation can be rewritten:
\begin{equation}
\Omega=\frac{\gamma_{\pm}^2m_{\text{Ca}^{2+}}m_{\text{S0}_4^{2-}}}
{(\gamma^{\text{sat}}_{\pm})^2m^{\text{sat}}_{\text{Ca}^{2+}}
m^{\text{sat}}_{\text{S0}_4^{2-}}}.
\label{omega}
\end{equation}
$m^{\text{sat}}_X$ is the molality of the aqueous component $X$ at chemical equilibrium,
and $\gamma^{\text{sat}}_{\pm}$ is the mean ionic activity coefficient at chemical
equilibrium.
Here we make the reasonable assumption that at all the ionic molalities achieved in our
experiments, the ionic strength remains small enough so that the mean ionic activity
coefficient does not depart from its value in pure water.
This term can thus be suppressed from the expression of Eq. \ref{omega}.

In the absence of experimental or theoretical determination of the value of
the constant $\mathcal{N}$ in Eq. \ref{trans}, we have used the value successfully used
for fitting calcite
dissolution measurements \citep{Rickard83,Shiraki}: $\mathcal{N}=1/2$.
This choice results in the dissolution flow rate at the solid-liquid interface:
\begin{equation}
F_{\text{dissolution}}(0,t)=
k s_r\left(1-\frac{\sqrt{m_{\text{Ca}^{2+}}(0,t)m_{\text{SO}_4^{2-}}(0,t)}}
{\sqrt{m^{\text{sat}}_{\text{Ca}^{2+}}m^{\text{sat}}_{\text{SO}_4^{2-}}}}\right).
\end{equation}
We will identify the geometric mean of the ionic molalities with our effective
molality $m$.
The geometric mean of the ionic solubilities is well known and will be written
$m^{\text{sat}}$.
Finally, the dissolution flow rate at the surface reduces to:
\begin{equation}
F_{\text{dissolution}}(0,t)=k s_r\left(1-\frac{m(0,t)}{m^{\text{sat}}}\right).
\label{disflowrate}
\end{equation}

The diffusion flow rate is linked to the diffusion flux at the
dissolving interface $J_{\text{diffusion}}(0,t)$ merely through:
\begin{equation}
J_{\text{diffusion}}(0,t)=\frac{1}{s}F_{\text{diffusion}}(0,t)
\label{difflowrate}
\end{equation}
where $s$ is the cell section perpendicular to the mass transport.
This flux derives readily from Fick's first law:
\begin{equation}
J_{\text{diffusion}}(0,t)=-D \rho\left(\frac{\partial m(0,t)}{\partial z}\right)_t
\label{fluxdif}
\end{equation}
with $\rho$ as the density of the solution.
The introduction of Eq. \ref{disflowrate}, \ref{difflowrate} and \ref{fluxdif} into
Eq. \ref{massbal} induces for the boundary condition at the lower end of the cell:
\begin{equation}
\left(\frac{\partial m(0,t)}{\partial z}\right)_t=
-\frac{k\beta}{D \rho}\left(1-\frac{m(0,t)}{m^{\text{sat}}}\right)
\label{condlim2}
\end{equation}
where we have introduced $\beta=s_r/s$.

\item At the upper end of the cell (water-air meniscus), the boundary
condition is merely written $J_{\text{diffusion}}(L,t)=0$, $L$ being the
height of water in the cell.

\end{itemize}

To get a tractable analytical expression of $m(z,t)$, we make a further
assumption concerning the geometry of our experiment.
Our cell is high enough and the investigated time
short enough to consider that gypsum diffuses in a semi-infinite medium.
This hypothesis has been experimentally verified, paying attention
to the fact that the highest molality fringe has not
reached the top of the cell at the end of the measurement.
The solution of Equation \ref{Fick2} is in this case \citep{Crank}:
\begin{equation}
m(z,t)=m^{\text{sat}}
\left[\text{erfc}\left(\frac{z}{2\sqrt{Dt}}\right)
-\exp\left[\frac{k\beta z}{D\rho m^{\text{sat}}}+(\frac{k\beta}{D\rho m^{\text{sat}}})^2Dt\right] 
\times\text{erfc}\left[\frac{z}{2\sqrt{Dt}}+\frac{k\beta}{D\rho
m^{\text{sat}}}\sqrt{Dt}\right]\right].
\label{semiinf}
\end{equation}

erfc is the complementary error function.
The values of $m^{\text{sat}}$, $D$ and $k$ are obtained through a fit of our experimental
$m(z,t)$ curves with the above expression.
The shift to the right of the molality curves (Fig. \ref{sscont}) can be viewed as a
signature
of diffusion, and the shift of the molality curves upward can be viewed as a signature of
dissolution.

\section{RESULTS}
\label{Results}

$\partial n/\partial m$ is a value of primary importance, but unfortunately no
experimental determination is available.
Therefore, we have 
carried out refractive index measurements with an Abbe refractometer
in pure water and water where gypsum was dissolved until saturation
at ambient temperature ($m^{\text{sat}}=15$ mmol kg$^{-1}$
corresponding to 2 g l$^{-1}$ \citep{Raju}).
These experiments were not far from the resolution limit of the
apparatus.
A (0.019 $\pm$ 0.007) kg mol$^{-1}$ value was found,
which was used for the 
refractive index into molality conversion.

To guarantee the validity of the data analysis , Equation \ref{Fick2} requires $D$
to remain constant with molality.
Unfortunately no experimental determination of $D$ exists, and furthermore of
the dependance of $D$ on molality.
To obtain an initial, crude idea of the variation $\Delta D$ of the diffusion
coefficient
of gypsum in water in the molality range $\Delta m$ between pure
water and saturated solution ($\Delta m=m^{\text{sat}}=15$ mmol kg$^{-1}$),
we have collected the $\partial D/\partial m$ values of other calcium
and sulfate salts in water \citep{Lobo,Zaytsev} and computed their $\Delta D$
for the same molality range.
The worst case is ZnSO$_4$ at low molality with $\Delta D/D\sim$ 20\%.
Evidently, no strict conclusion can be inferred for gypsum from that value,
but a relatively weak dispersion of $D$ in our experimental cell
can be expected.

Precipitation and dissolution of minerals are known to be strongly influenced by
the presence of impurities.
To evaluate the purity of our single crystals, we carried out the Electron Probe Microanalysis
of a sample of our material. 
No elements, other than the atoms constituting gypsum, have been detected with a concentration
greater than 0.1\%.
The known impurities found in the gypsum of the Mazan mines, possibly present with a
lower concentration, are dolomite rocks, quartz, anhydrite and celestite.

For the computation of $\beta=s_r/s$ used in Equation \ref{semiinf},
the cross-diffusional section area $s$
is taken as the horizontal section of our optical cell ($s$ = 0.95 cm$^2$)
and $s_r$ is taken as the dissolving surface area of the single crystal ($s_r\approx$
0.3 cm$^2$).

We have measured the $pH$ of our solution at the end of the experiment and
observe a pH of 5.7, which roughly corresponds to the $pH$ of the
CO$_2$/HCO$_3^-$ system stemming from the dissolution of the atmospheric CO$_2$ in water.

Seven measurements, all at 20.00$^\circ$C, have been performed to check the reproducibility.
The fit of the experimental points by the theoretical law of Eq. \ref{semiinf}
brings a solubility $m^{\text{sat}}=15.6$ mmol kg$^{-1}$, which is very close to the expected
value of 15 mmol kg$^{-1}$ \citep{Raju}.
This agreement validates our analysis procedure and particularly the chosen
$\partial n/\partial m$ value.
We find a diffusion coefficient
$D=(1.0\:\pm\: 0.1)\times10^{-9}$ m$^2$ s$^{-1}$,
a value very close to the $0.9\times10^{-9}$ m$^2$ s$^{-1}$
value derived from Nernst-Hartley equation,
generally used in dissolution studies \citep{Barton}.
We find a dissolution rate constant $k=(4\:\pm\: 1)\times10^{-5}$
mol m$^{-2}$ s$^{-1}$.
The statistical uncertainty is the standard error, computed from the
seven measurements.

\section{MIXED KINETICS}
\label{mixed}

There has been a strong debate to clearly determine the nature and the kinetics
of the slowest step of the dissolution process, controlling the kinetics.
If mass transport in the liquid (proportional to the diffusion coefficient $D$)
is slow compared to the chemical reaction at the solid surface (proportional
to the dissolution rate constant $k$), ions are rapidly unbound from the solid but slowly
transported away in the solution, hence the dissolution global kinetics is controlled by
molecular diffusion.
Conversely, if mass transport proceeds quickly compared to the reaction rate, the kinetics is
controlled by the slow ion detachment from the mineral.
Both macroscopic studies ---rotating-disk, column-experiments \dots
\citep{Raines,Dreybrodt,Dewers,Jeschke}--- and microscopic studies ---
atomic force
microscopy \citep{Bosbach,Hall}--- have tried to address the question.

One of the reasons of the diversity of the available results lies in the
fact that the balance between transport and reaction strongly depends on the geometry
of the system and on the thermodynamic conditions.
\cite{Jeschke} conclude with mixed kinetics,
where both effects are of
comparable magnitude, with the linear kinetics of Eq. \ref{disflowrate}
($F_{\text{dissolution}}\sim m$, except very close to the saturation) for the dissolution.
This conclusion is drawn from a combination of batch dissolution
and rotating-disk experiments.
Therefore, the authors make use of the formalism of these kinds of methods to
evaluate the two phenomena: the dissolution velocity is evaluated by $k_s$,
an empirical dissolution rate constant, and the diffusion velocity is evaluated by
$k_t=Dc_{\text{eq}}/\varepsilon$, a transport constant ($c_{\text{eq}}$ stands for the
molarity of aqueous calcium at saturation and $\varepsilon$ for the
thickness of the diffusion boundary layer surrounding the crystal).
The authors find $k_t=1.5\times 10^{-3}$ mol m$^{-2}$ s$^{-1}$ from batch experiments and
$k_s=1.3\times 10^{-3}$ mol m$^{-2}$ s$^{-1}$ from both experiments
(for $c/c_\text{eq}<0.94$).
These two values are so close that the mixed aspect of the dissolution kinetics is
unambiguously assessed for this situation.
Beyond the interest of this result, one can see that the selected evaluation quantities
are strongly linked to the experimental methods.

Besides,
\cite{Murphy} have proposed the ratio of the maximum diffusive and reactive
flow rates $\alpha=D\rho m^{\text{sat}}/(\beta k L)$ to be a dimensionless number used to
discriminate between transport-controlled and reaction-controlled dissolution.
This number has a more universal impact, with no quantity intrinsic to the
chosen experiments (like $\varepsilon$ above).
Pure transport control corresponds theoretically
to $\alpha=0$, pure surface control to
$\alpha=\infty$ and mixed control to $\alpha=1$.
These authors have numerically investigated realistic geological
dissolution configurations of quartz and calcite in water in geometries comparable to ours
(zero-flux boundary) and
conclude that at values of $\alpha$ between $10^{-2}$ and $10$, the rate of evolution
of the system is controlled by mixed surface reaction and diffusion kinetics.

Our experimental values lead, with $L=3$ cm and $\rho=10^3$ kg m$^{-3}$,
to $\alpha=0.05$.
This reveals mixed kinetics, according to the classification of Murphy et al.
At ambient temperature and pressure, the mixed kinetics of dissolution
of gypsum in water seems to be a robust feature, in our geometry (a reacting surface
at one boundary and a zero flux at the other) as well as in the
geometry of \cite{Jeschke} (a reacting surface at one boundary
and a constant composition reservoir at the other).

\section{COMPARISON BETWEEN LOCAL AND GLOBAL DISSOLUTION MEASUREMENTS}

The comparison of our dissolution rate constant $k$ values with results
of global measurements requires particular care.
As has been often stated \citep{Rickard83,Jeschke02}, derivation of rate
constants from these kinds of experiments is a difficult task and the procedure
strongly depends on the device geometry and sample morphology.

To enter into the details of a global experiment,
the overall flux $\mathcal{R}$ in a vessel of volume $v$ is computed
from the time evolution of the
overall concentration $c$ and is considered to be identically
equal to the diffusion and dissolution fluxes:
$\mathcal{R}=J_{\text{dissolution}}=J_{\text{diffusion}}=(v/s)(dc/dt)$.
But as water is flowing in the vessel, this equality is only valid
at the frontier between the diffusional boundary layer
and the bulk liquid, in other words along the section called $s$ above.
Indeed this section corresponds to the geometrical locus where mass balance
of Eq. \ref{massbal} applies.
This statement remains true exclusively for mixed kinetics.
For pure reaction-controlled dissolution, mass balance applies at the solid-liquid
interface, called $s_r$ above, and for pure transport-controlled dissolution, no
information on dissolution coefficients can be obtained \citep{Jeschke02b}.

Between the solid surface and the extremity of the boundary layer, there is no loss
of solute, therefore mass balance imposes the flow rates at these two surfaces to equalize.
At the solid-liquid interface (of area $s_r$),
the flow rate $F_r$ can be deduced from an analog to Eq. \ref{disflowrate}:
$F_r=s_rk(1-c/c_{\text{eq}})$.
At the top of the diffusional boundary layer (of area $s$), the flow rate can be
written $F_d=sk_s(1-c/c_{\text{eq}})$ where $k_s$ is the
empirical dissolution rate constant introduced in Section \ref{mixed}
\citep{Jeschke}.
Accordingly $F_r=F_d$ implies $k=sk_s/s_r$ and we are now able to compare our microscopic
dissolution rate constant $k$ with a rate constant $k_s$ deduced from a global dissolution
measurement.

Recently, \cite{Jeschke} found
a linear dissolution kinetics and $k_s=1.3\times 10^{-3}$ mol m$^{-2}$ s$^{-1}$,
as previously mentionned.
The authors estimate the geometric specific surface area of their
powders by optical microscopy to range between 60 and 73 cm$^2$ g$^{-1}$, depending on the gypsum type.
We consider this surface to be similar to our diffusional section $s$.
The BET-surface of these powders is 1100 cm$^2$ g$^{-1}$.
We identify this surface with our dissolving interface $s_r$
\footnote{Obviously this is questionable because the adsorption sites in the BET method
and the dissolving sites in the dissolution experiment may slightly differ.
But this assumption should at least bring the correct order of magnitude.}.
Therefore,
one finds a $s/s_r\sim 1/15$ ratio, which gives a $k=9\times10^{-5}$
mol m$^{-2}$ s$^{-1}$ value.

Beside the uncertainty on the surface ratio, a factor of 2 or 3 between their
global measurements ($9\times10^{-5}$
mol m$^{-2}$ s$^{-1}$) and our local measurements ($4\times10^{-5}$ mol m$^{-2}$ s$^{-1}$)
seems to exist
at first sight.
But the large differences between the two
methodologies could account for this discrepancy and we can consider these two
values as being in fair agreement.
One should mention that the $k$ we have measured concerns exclusively
the dissolution
of the (010) plane, whereas the $k_s$ of global experiments is an average of the
dissolution rate constants of all the reacting planes of the dissolving powder.
Therefore, the two values are not strictly comparable.

As a summary, microscopic measurements make the data interpretation easier for three
reasons.
First, there is no uncertainty on the surface where mass balance must be applied.
Second, there is no flow in the cell and therefore no hydrodynamical assumptions and
computations are needed.
And finally the study of a single crystallographic interface avoid to obtain
a multifaceted average dissolution rate constant.

\section{CONCLUSION}

We have used a non-invasive interferometric method to access to the microscopic
dissolution rate constant of a cleaved (010) surface of gypsum in water.
This constant has been revealed to be two times lower than
the same constant measured by macroscopic averaged methods,
which can be considered to be in fair agreement.
The interdiffusion coefficient of the aqueous species of
gypsum in water has also been measured
during these experiments and exhibits a value close to the 
theoretical value generally used for dissolution studies.
Now, to deepen the comprehension of pressure solution creep, we plan to
perform the same experiments with gypsum crystals under stress.
The knowledge of the influence of an uniaxial pressure on the dissolution
parameters may shed new light on the mechanisms of PSC.

\vspace{3ex}

\newpage

\small
\noindent
\textit{Acknowledgements---} We would like to thank Elisabeth Charlaix and Pierre Monchoux
for fruitful discussions, Sylvain Meille (LCR Lafarge) for gypsum samples,
Xavier Jaurand for the chemical analysis of gypsum,
and Christophe Bineau, Alexandre Enderlin, Herv\'e F\'eret and Agn\`es Piednoir
for experimental help.
We are also very grateful to the anonymous referees and to the associate
editor, Eric H. Oelkers, for their comments.
Part of this study was funded by CNES (French spatial agency).

\newpage

\listoftables

\renewcommand{\baselinestretch}{1}

\begin{table}
\begin{center}
\begin{tabular}{c c c}
\hline
Symbol         & quantity                                       & units \\
\hline
$\beta$        & geometrical constant ($s_r/s$)                  & \\
$\gamma_{\pm}$ & mean ionic activity coefficient                 & \\
$\gamma^{\text{sat}}_{\pm}$ & mean ionic activity coefficient at equilibrium & \\
$\Delta\phi$   & phase difference                                & \\
$\Delta D$     & diffusion coefficient difference                & m$^2$ s$^{-1}$ \\
$\Delta m$     & molality difference                             & mol kg$^{-3}$ \\
$\Delta n$     & refractive index difference                     & \\
$\varepsilon$  & thickness of a diffusion boundary layer         & m \\
$\lambda$      & laser wavelength                                & m \\
$\xi$          & advancement variable                            & mol \\
$\rho$         & density of the solution                         & kg m$^{-3}$ \\
$\Omega$       & supersaturation                                 & \\
$A$            & chemical affinity of dissolution                & J mol$^{-1}$ \\
$a_{\text{X}}$ & activity of species X                           & \\
$a^{\text{sat}}_{\text{X}}$ & activity of species X at equilibrium & \\
$c$            & molarity of aqueous calcium                     & mol m$^{-3}$ \\
$c_{eq}$       & molarity of aqueous calcium at equilibrium      & mol m$^{-3}$ \\
$D$            & interdiffusion coefficient of the solution      & m$^2$ s$^{-1}$ \\
$e$            & path length in the optical cell                 & m \\
$F$            & flow rate                                       & mol s$^{-1}$ \\
$J$            & flux                                            & mol m$^{-2}$ s$^{-1}$ \\
$k$            & dissolution rate constant                       & mol m$^{-2}$ s$^{-1}$ \\
$k_s$          & empirical dissolution rate constant             & mol m$^{-2}$ s$^{-1}$ \\
$k_t$          & transport constant ($=Dc_{eq}/\varepsilon)$      & mol m$^{-2}$ s$^{-1}$ \\
$K_{sp}$       & solubility product                              & \\
$L$            & height of water                                 & m \\
$m$            & effective molality of dissolved gypsum          & mol kg$^{-1}$ \\
$m^{\text{sat}}$ & effective molality of dissolved gypsum at equilibrium & mol kg$^{-1}$ \\
$m_{\text{X}}$ & molality of species X                           & mol kg$^{-1}$ \\
$m^{\text{sat}}_{\text{X}}$ & molality of species X at equilibrium & mol kg$^{-1}$ \\
$n$            & index of refraction                             & \\
$N$            & number of interference fringes                  & \\
$\mathcal{N}$     & numerical constant                              & \\
$R$            & gas constant                                    & J mol$^{-1}$ K$^{-1}$ \\
$\mathcal{R}$  & overall flux                                    & mol m$^{-2}$ s$^{-1}$ \\
$s$            & area of the section perpendicular to mass transport & m$^2$ \\
$s_r$          & area of the dissolving interface                & m$^2$ \\
$T$            & absolute temperature                            & K \\
$t$            & time                                            & s \\
$z$            & vertical coordinate                             & m \\
\hline
\end{tabular}
\label{symb}
\caption[Description of the symbols used.]{}
\end{center}
\end{table}

\clearpage

\renewcommand{\baselinestretch}{2}

\listoffigures

\newpage

\clearpage

\clearpage

\end{document}